\documentclass[twocolumn,english,aps,prl,amsmath,amssymb]{revtex4}

\usepackage{dcolumn}
\usepackage{mathrsfs}
\usepackage{breakcites}
\usepackage{latexsym}
\usepackage{hyperref}
\usepackage{bm}
\usepackage[T1]{fontenc}
\usepackage[utf8]{inputenc}
\usepackage{ae}
\usepackage{graphicx}
\usepackage{epsfig}
\usepackage{pstricks,pst-grad}
\usepackage{babel}
\usepackage{hyperref}
\usepackage{ulem}
\usepackage{latexsym}
\usepackage{amsfonts}
\usepackage{amssymb}
\usepackage{amsmath}
\usepackage{color}
\usepackage{verbatim}
\usepackage[justification=raggedright]{caption,subfig}
\usepackage{xr}

\newcommand{\Nbar}{{\bar{\cal N}}}
\newcommand{\R}{{R_{{\rm e}0}}}
\newcommand{\ie}{{\it i.e.}}

\newcommand{\br}{{\bf r}}
\newcommand{\bc}{{\bf c}}

\newcommand{\rd}{{\rm d}}

\newcommand{\kT}{k_{\rm B}T}

\hypersetup{
    	pdftitle={note} %ADAPT TEXT 
        pdfauthor={Marcus M&uuml;ller},  %ADAPT TEXT 
	pdfsubject={},
	pdfproducer={},
	pdfview=FitV,       % FitH
	pdfstartview=FitB,
	linkcolor=blue,     % for links on same page
	citecolor=blue,     % for citations
	urlcolor=red,      % for Links to URLs
	breaklinks=true,    % Links can be be line-broken
	colorlinks=true,
	citebordercolor=0 0 0,  % color for \cite
	filebordercolor=0 0 0,
	linkbordercolor=0 0 0,
	menubordercolor=0 0 0,
	urlbordercolor=0 0 0,
	pdfhighlight=/I,
	pdfborder=0 0 0,   % no Box around links
	bookmarksopen=true,
	bookmarksnumbered=true
}

\begin{document}
 
\title{Interface repulsion and lamellar structures in thin films\\
of homopolymer blends due to thermal oscillations}

\author{Louis Pigard}
\author{Marcus M\"uller}
\email{mmueller@theorie.physik.uni-goettingen.de}
\affiliation{University of Goettingen, Institute for Theoretical Physics, 37077 G{\"o}ttingen, Germany}
\begin{abstract}
In equilibrium the interface potential that describes the interaction between two AB interfaces in a binary blend of A and B homopolymers is attractive at all distances, resulting in coarsening of the blend morphology even in the absence of interface curvature. We demonstrate that the dissipative assembly in response to a time-periodic variation of the blend incompatibility qualitatively alters this behavior, i.e., for suitable parameters the interface potential exhibits a periodic spatial modulation and AB interfaces adopt a well-defined distance. We explore for which oscillation periods and amplitudes an interface repulsion occurs and demonstrate that we can control the preferred interface distance over a wide range by varying the oscillation period. Using particle-based simulations we explicitly demonstrate that this dissipative assembly of a homopolymer blend results in a lamellar structure with multiple planar interfaces in a thin film geometry.
\end{abstract}

\maketitle
Dissipative assembly, i.e., the non-equilibrium structure formation due to energy input and concomitant dissipation, is an important mechanisms of structure formation in biological and synthetic systems \cite{Fialkowski2006,Seeley2002,Whitesides2002}. Recently there are efforts to mimic particular aspects of this driven, non-equilibrium assembly in colloidal systems \cite{Igal1, Igal2, Swan12,Swan16,Swan18}, and there is an urgent need to understand the underlying design mechanisms \cite{Fialkowski2006,Igal1}. Here we provide an analytical strategy for predicting the dissipative assembly of interfaces and demonstrate that the non-equilibrium structure formation results in a wetting behavior that qualitatively differs from the equilibrium behavior, i.e., the interface interactions switch from attractive to repulsive, giving rise to the dissipative assembly into lamellar morphologies.

The interaction between interfaces is important for wetting and coarsening dynamics \cite{cahn58,Hohenberg77,deGennes1985,Bray94,Binder03}. The free energy of two planar interfaces per unit area depends on their distance, $h$, and is given by the interface potential, $g(h)$ \cite{schick1990introduction}. In equilibrium, the interaction between two interfaces in a binary blend is attractive \cite{Kawasaki82}, resulting in coarsening even in the absence of interface curvature or collision of interfaces due to thermal motion. In this letter we demonstrate that an oscillatory repulsion between the species qualitatively changes the interface potential, resulting in an interface repulsion and giving rise to a metastable lamellar morphology in a binary blend.

The universal properties of interfaces between the coexisting phases of a binary blend can be described by the Ginzburg-Landau square-gradient theory \cite{cahn58,binder84nuc}: 
\begin{equation}
\frac{{\cal F}[m]}{\kT{\cal N}} =  \int \rd \br\; \left( -\frac{\alpha}{2}m^{2}+\frac{1}{4}m^{4}+\frac{1}{2} \left| \nabla m\right|^{2}\right)
\label{eqn:GL}
\end{equation}
where $m$ denotes the order parameter that distinguishes between the two coexisting phases, and $\cal N$ is a normalization that sets the scale of the free-energy density in units of the thermal energy, $\kT$. The coefficient $\alpha$ parameterizes the strength of the binary interaction between the constituents of the blend. It is a temperature-like variable that controls the phase behavior, and $\alpha=0$ corresponds to the critical point. The square-gradient term penalizes rapid spatial variations of the order parameter and results in a finite width of the interface. The interface profile 
$
m(x) = \sqrt{\alpha}\tanh(x/w)
$
with width $w=\sqrt{2/\alpha}$ minimizes the free-energy functional, ${\cal F}[m]$, subjected to the boundary conditions $m(x\to\pm\infty)=\pm \sqrt{\alpha}$ \cite{cahn58}. The interaction between two interfaces is attractive at all distances and decays exponentially with the distance, $h$. Within Cahn-Hilliard dynamics
\begin{equation}
\frac{\partial m}{\partial t} = \nabla \frac{\Lambda}{\kT {\cal N}} \nabla \frac{\delta {\cal F}}{\delta m(\br)}
\label{eqn:CH}
\end{equation}
this yields to an attractive dynamics of the interface distance,
$
h \frac{\rd h}{\rd t} \sim - \exp(-h/w)
$
for $\alpha=1$ \cite{Kawasaki82}.

In this letter we consider the case that the incompatibility, $\alpha$, is a time-periodic function 
$
\alpha = \left\{ 
\begin{array}{ll}
1+\epsilon & \mbox{for}\; 0<t \leq T/2 \\
1-\epsilon & \mbox{for}\; T/2<t \leq T
\end{array}\right.
$
with period, $T$, and amplitude, $\epsilon$, of the oscillation. Previous seminal work by Szleifer and co-workers \cite{Igal1,Igal2} for particle systems demonstrated that oscillations of the interparticle potential with a period that is fast compared to the diffusion timescale of the particles result in an effective interaction that is the time average of the oscillatory potential. The (slow) toggling of interactions also beneficially impacts the crystallization kinetics in colloidal systems \cite{Swan12,Swan16,Swan18}.

In our continuum  model, we demonstrate by numerical simulation of the Cahn-Hilliard dynamics, approximate analytical calculations, and simulation of a particle-based model of a symmetric homopolymer blend in a thin film, that an oscillatory incompatibility can qualitatively change the interface potential from attractive to repulsive -- a non-equilibrium characteristic that cannot be observed for any averaged (time-independent) value of $\alpha$. The continued input of energy in this time-periodic non-equilibrium state results in a preferred distance between interfaces and a dissipative assembly of a binary blend into a metastable lamellar configuration, whose lamellar spacing can be controlled by the period, $T$, of the oscillation of the strength of binary interactions.

\begin{figure}[t]
\includegraphics[width=0.9\columnwidth]{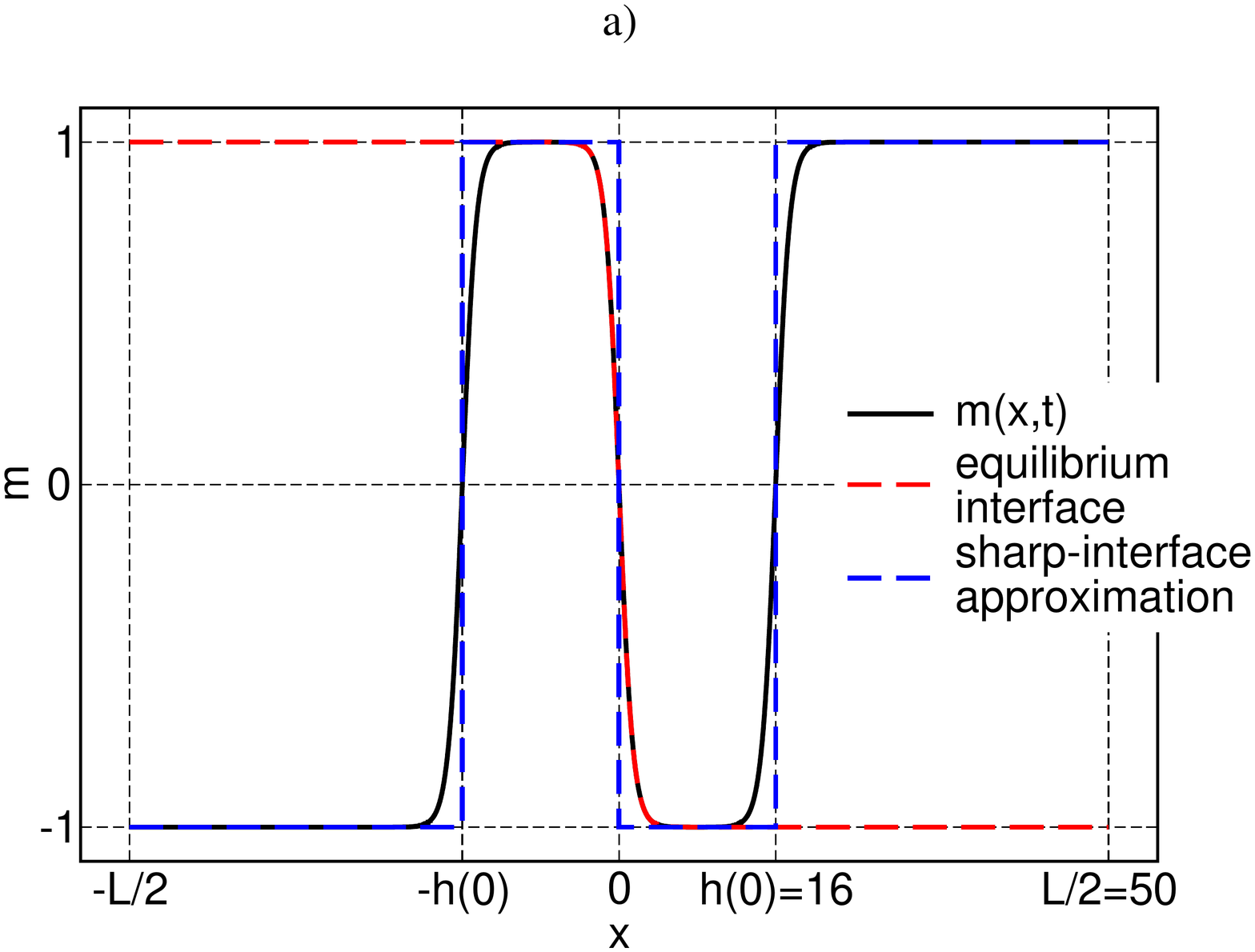}
\includegraphics[width=0.9\columnwidth]{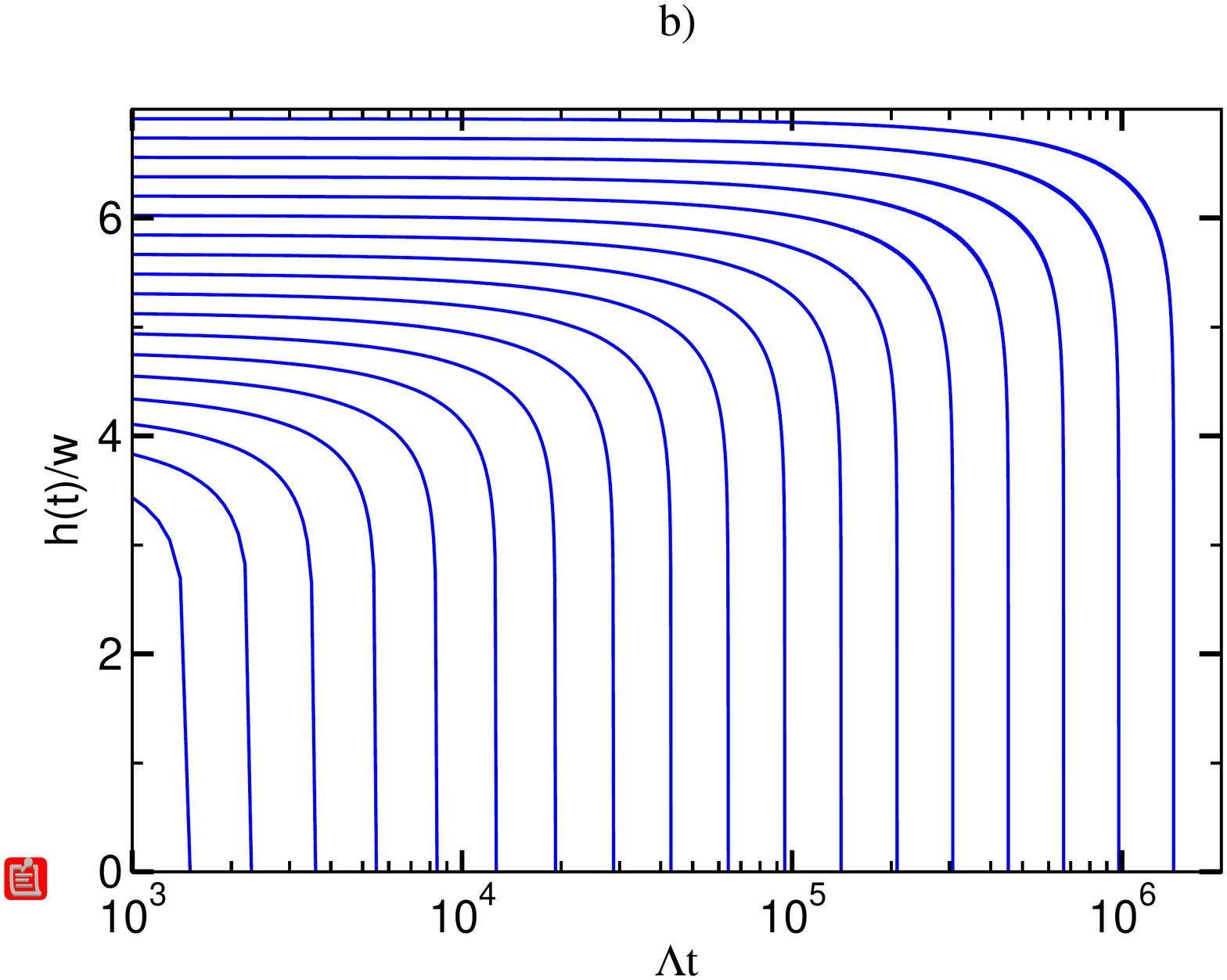}
\caption{a) Initial interface configuration with an interface distance, $h(0)=16\approx11.3w$, comparison with the equilibrium interface profile for $\alpha \equiv 1$ and the sharp-interface approximation. b) Time evolution towards equilibrium for a constant incompatibility, $\alpha \equiv 1$, \ie, $\epsilon=0$: interfaces attract each other, their distance $h(t)$ decreases in time, and they annihilate at $h(t^*)=0$. The concomitant time scale, $t^*$, increases exponentially with the initial distance, $h(0)$.
}
\label{fig:1}
\end{figure}

First, we study the Cahn-Hilliard dynamics of the Ginzburg-Landau square-gradient functional via a semi-implicit spectral technique, setting ${\Lambda}=1$. We consider a one-dimensional system of size $L=100$. Initially, three interfaces are symmetrically positioned at $-h(0)$, 0, and $h(0)$ as shown in \autoref{fig:1}a. Panel b shows the time evolution of the distance, $h(t)$, for constant $\alpha\equiv1$, \ie, $\epsilon=0$. As expected \cite{Kawasaki82} the interfaces attract each other and annihilate at a time $\ln \Lambda t^{*} \sim h(0)$ for $h(0)\gg1$.

\begin{figure}[t]
\includegraphics[width=0.9\columnwidth]{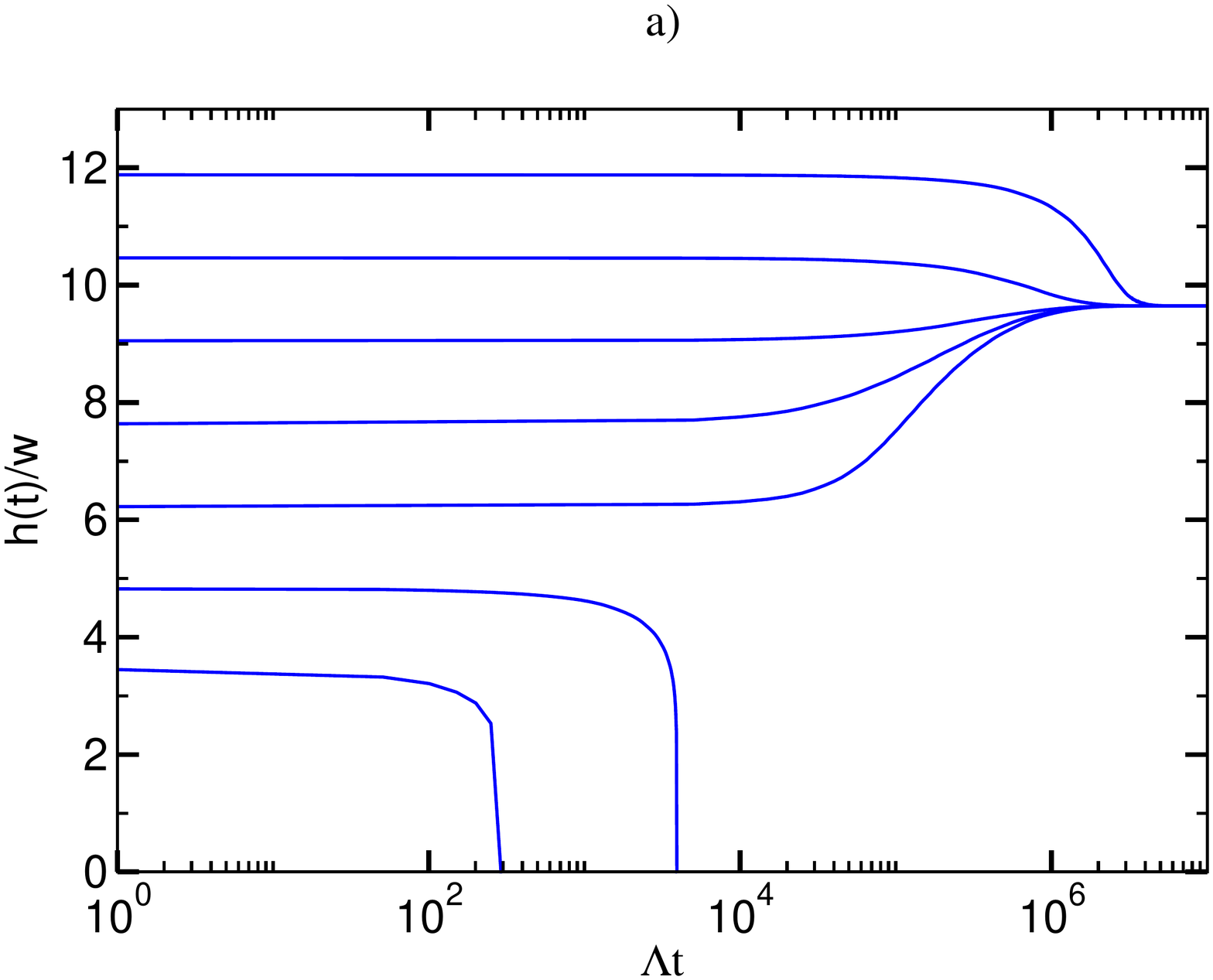}
\includegraphics[width=0.9\columnwidth]{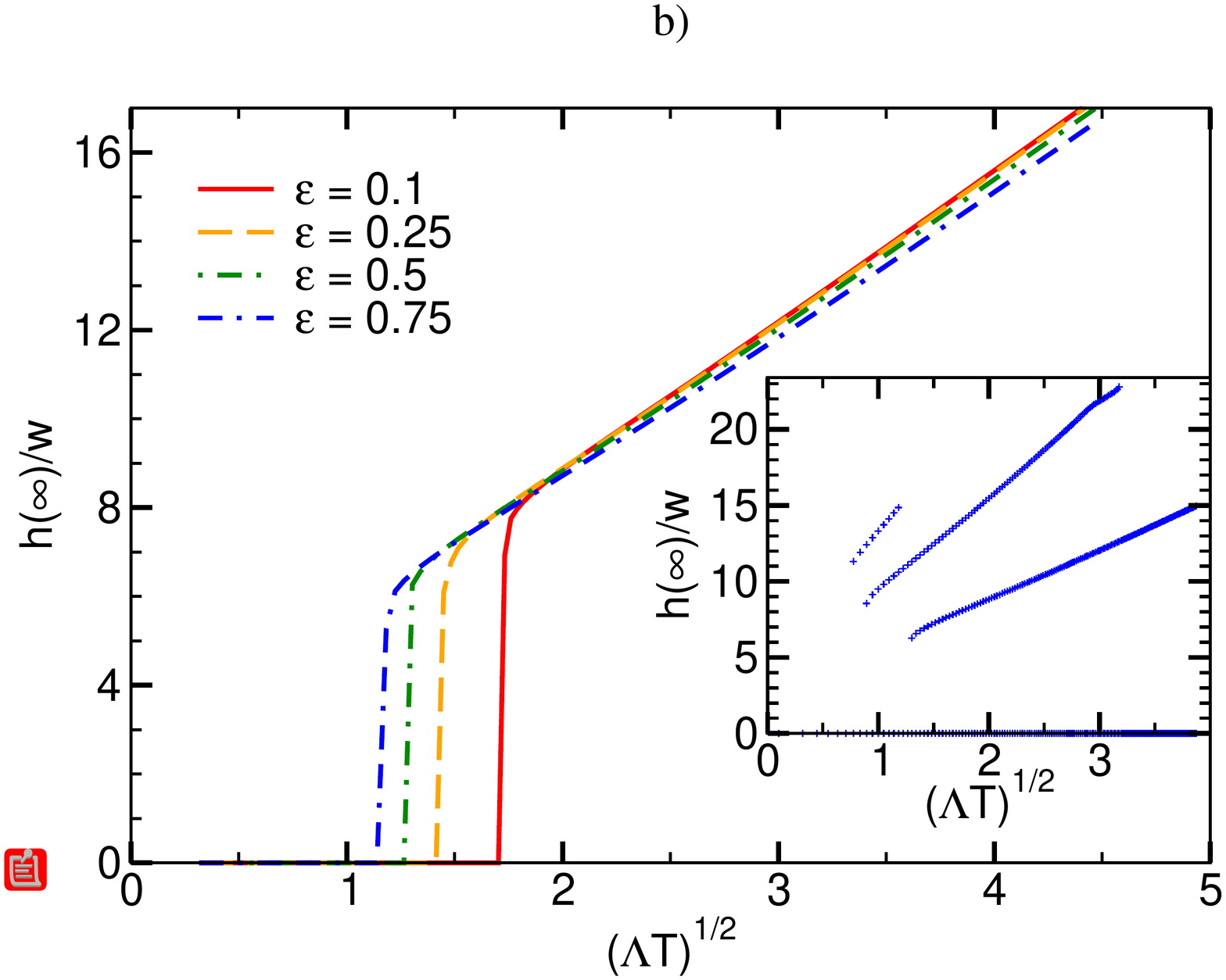}
\caption{
a) Time evolution of the interface distance, $h(t)$, for different starting conditions, $h(0)$, for an oscillatory incompatibility with amplitude $\epsilon=0.5$ and period $\Lambda T=5$.
b) $h(\infty)$ as a function of the square-root of the oscillation period, $T$ for various amplitudes, $\epsilon$. The inset illustrates further fix-point distances for $\epsilon=0.5$
}
\label{fig:2}
\end{figure}

The behavior for oscillating $\alpha(t)$ with amplitude $\epsilon=0.5$ and period $\Lambda T=5$ is presented in \autoref{fig:2}a. If the interfaces are initially closely spaced, $h(0) \lesssim 6 w$, they also attract each other and annihilate. For larger initial distances, however, interfaces do not attract. Instead, the interfaces converge toward a well defined, finite separation distance, $h(\infty) \approx 9.6 w$.

\autoref{fig:2}b shows how the fix-point distance, $h(\infty)$, depends on the amplitude, $\epsilon$, and period, $\Lambda T$ of the oscillation. For very fast oscillations, $T<T^*$, interfaces attract each other and annihilate, \ie, there is no non-trivial fix-point distance, $h(\infty)>0$; in accord with the prediction of Szleifer and co-workers the driven system behaves similar to an equilibrium system with time-averaged interactions \cite{Igal1,Igal2}. For slower oscillations, however, there emerges a preferred distance, $h(\infty)$, between the interfaces, and it increases with the period of the oscillations of incompatibility. The amplitude of the oscillations, $\epsilon$, slightly shifts the threshold period, $T^*$, to smaller values but has only a minor influence on $h(\infty)$ for larger periods. The inset of \autoref{fig:2}b illustrates that there exist multiple preferred distances for a fixed oscillation period $T$; in the following we focus on the smallest non-trivial $h(\infty)$, observed in the numerical Cahn-Hilliard dynamics.

\begin{figure}[t]
\includegraphics[width=0.9\columnwidth]{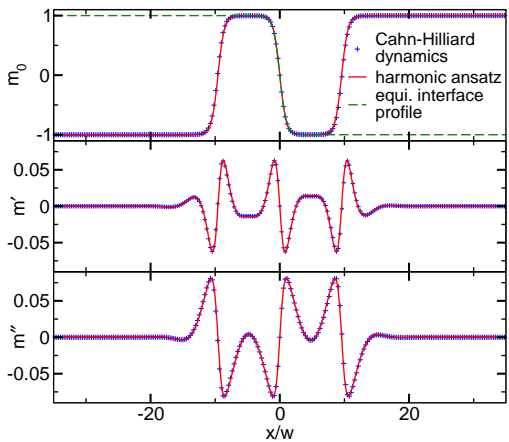}
\caption{
Harmonic analysis of the profile after the fix-point has been approached for $\epsilon=0.5$ and $\Lambda T=5$. The upper-panel presents the period-averaged profile, $m_0(x)=\lim_{t\to \infty}\bar m(x,t)$, whereas the lower panels show the in-phase and out-of-phase oscillations. Crosses correspond to the results of the numerical Cahn-Hilliard dynamics whereas the solid lines are the solution of \autoref{eqn:QPA}.
}
\label{fig:3}
\end{figure}

To analyze these findings, we decompose the time-dependent order-parameter profile,  $m(x,t)$, in the form
\begin{equation}
m(x,t) \approx m_{0}(x) + \epsilon \left[ m_{1}(x) e^{i \omega \Lambda t} + \mathrm{c.c.} \right] \; \mbox{with}\;  \omega=\frac{2\pi}{\Lambda T}
\label{eqn:ansatz}
\end{equation}
after the fix-point distance has been approached. Higher harmonics are ignored. $m_{0}$ is the time-averaged profile for $t \to \infty$, and $m'=-2\epsilon \Im(m_{1})$ and $m''=2\epsilon \Re(m_{1})$ denote the in-phase and out-of-phase oscillations, respectively. \autoref{fig:3} shows that $m_{0}$ closely resembles the equilibrium profile with the time-averaged $\bar \alpha = 1$, whereas the in-phase and out-of-phase contributions exhibit an additional spatial structure that dictates $h(\infty)$. 

To make progress, we insert this harmonic ansatz (\ref{eqn:ansatz}) into the kinetic equation (\ref{eqn:CH}), approximate $\alpha(t) \approx1+\frac{4\epsilon}{\pi}\sin \omega \Lambda t + \cdots$, and obtain to leading orders in multiples of the frequency $\omega$
\begin{eqnarray}
0 &=& - m_{0} + m_{0}^{3} -\nabla^{2}m_{0} + \epsilon^{2} \left[ \frac{4}{\pi}\Im(m_{1})+6 m_{0} |m_{1}^{2}|\right] \nonumber \\
i \omega m_{1} &=&  \nabla^{2} \left( \frac{2i}{\pi} m_{0} + (3m_{0}^{2}-1)m_{1} - \nabla^{2}m_{1}\right) \label{eqn:QPA}
\end{eqnarray}
where we have twice integrated the first equation with respect to $x$, using zero-flux boundary conditions and $m(-x) = -m(x)$. The numerical solution of these equations is also shown in \autoref{fig:3} as lines, indicating excellent agreement between our ansatz (\ref{eqn:ansatz}) and the long-time, numerical solution of the Cahn-Hilliard dynamics.

In \autoref{fig:2} we observe that the convergence of the interface distance, $h(t) \to h(\infty)$, is much slower than an oscillation period, $T$. We exploit this time-scale separation by assuming that the profile, $\bar m(x,t) \equiv \frac{1}{T} \int_{t}^{t+T} \rd t'\; m(x,t')$, averaged over one period is dictated by the slow dynamics of $h(t)$, whereas the oscillatory part of the profile rapidly adjusts to $\bar m(x,t)$. Thus, we can use \autoref{eqn:QPA} with $m_{0}$ replaced by $\bar m$ to obtain $m_{1}$ as a functional of $\bar m$.

Inserting this functional $m_{1}[\bar m]$ into the Cahn-Hilliard equation (\ref{eqn:CH}), we obtain an approximation for the slow dynamics of $\bar m(x,t)$
\begin{eqnarray}
\frac{\partial \bar m}{\partial t} &=&  \nabla \frac{\Lambda}{\kT {\cal N}} \nabla \left( \bar \mu + \epsilon^{2}  \mu_{2} \right)\\
\mbox{with}\quad \frac{\bar \mu}{\kT {\cal N}} &=&  \frac{1}{\kT {\cal N}}\frac{\delta {\cal \bar F}}{\delta \bar m} =  - \bar m + \bar m^{3} -\nabla^{2}\bar m \\
\frac{\mu_{2}}{\kT {\cal N}} &=&  \frac{4}{\pi}\Im(m_{1}[\bar m])+6 \bar m |m_{1}[\bar m]|^{2} \label{eqn:mu2}
\end{eqnarray}
where ${\cal \bar F}$ is the original Ginzburg-Landau free-energy functional (\ref{eqn:GL}) with the time-averaged incompatibility $\bar \alpha=1$.

In the Supplementary Information (SI) we show that we can approximate the additional term as the derivative of a Lyapunov functional,
$
\mu_{2} \approx \frac{\delta {\cal L}_{2}}{\delta \bar m} 
$, with
\begin{equation}
\frac{{\cal L}_{2}}{\kT{\cal N}} = -\frac{1}{2} \int \rd x \rd y \; \nabla \bar m(x) K(x-y) \nabla \bar m(y)\,. \label{eqn:L2}
\end{equation}
Thus, the time evolution of the slow period-averaged profile, $\bar m(x,t)$,  approximately follows a Cahn-Hilliard dynamics with the effective free-energy functional
$
{\cal F}_{\rm eff}[\bar m]={\cal \bar F}[\bar m]+\epsilon^{2}{\cal L}_{2}[\bar m]
$. In particular, fix-points of the oscillatory dynamics correspond to extrema of ${\cal F}_{\rm eff}[\bar m]$. Using a sharp-interface approximation for $\bar m(x,t)$, we obtain (see SI)
\begin{eqnarray}
F_{\rm eff}(h) &=& {\cal F}_{\rm eff}[\bar m] = 2g(h)-g(2h) \\
\mbox{with}&& g(h) = 4 \kT {\cal N}\epsilon^{2} K(|h|) + \mbox{const}
\label{eqn:Feff}
\end{eqnarray}

The kernel, $K$, in \autoref{eqn:L2} is the sum of two damped, spatially periodic modulations  (see SI)
\begin{eqnarray}
K(h) = \Re\left(c_{+}e^{-\lambda_{+}|h|} + c_{-}e^{-\lambda_{-}|h|}\right)  \quad \mbox{with} \\
\lambda_{\pm}^2 = 1\pm\sqrt{1-i\omega} \quad \mbox{and} \quad
c_{\pm} = \frac{2}{\pi^{2} \lambda_{\pm}(1- \lambda_{\pm}^{2})} \label{eqn:res}
\end{eqnarray}
For slow oscillations, $\omega \ll 1$, we obtain $\lambda_{+}\approx 2$ and $\lambda_{-}\approx \sqrt{i\omega/2}$. Since $1/\Im(\lambda_{-})$ sets the characteristic length scale of the spatial modulation of $K$, this approximation suggest that the fix-point distance scales like $h(\infty) \sim \omega^{-1/2} \sim \sqrt{\Lambda T}$, in qualitative agreement with the numerical data in \autoref{fig:2}. Additional information about the energy dissipation is provided in the SI.

Lastly, we illustrate the predictions of the Ginzburg-Landau model by particle-based simulations of a symmetric binary polymer blend. Binary polymer blends are well described by the Ginzburg-Landau free-energy functional \cite{binder84nuc,Reister01,Muller05f}. Importantly, the parameter $\alpha$ of the Ginzburg-Landau model is related to the distance, $\chi N -2$, to the critical point of demixing, where $N$ denotes the number of segments per polymer and $\chi$ the Flory-Huggins parameter. Thus, minor changes in the pairwise interactions, $\chi$, between segments give rise to pronounced changes in the phase diagram. This sensitivity of the phase behavior of a polymer blend stems from the small translational entropy of the macromolecules and has also been exploited previously in the context of active polymer systems \cite{Kactive}. Moreover, recent experiments by Kriisa and Roth \cite{Roth19} have realized periodic jumps from the one-phase region of the phase diagram into the miscibility gap in polymer blends by subjecting a blend to an oscillating electric field. Additionally, the dynamics of polymers is slow in comparison to that of mixtures of low-molecular weight components and the predicted repulsion between interfaces occurs for moderate and slow oscillations. Thus binary polymer blends are a promising model system to validate our predictions.

The simulations use a soft, coarse-grained model \cite{mm11b} in conjunction with the single-chain-in-mean-field algorithm \cite{Daoulas06b} implemented in the SOMA program \cite{SOMA}. The system of geometry $12 \times 2 \times 2 \R^{3}$, where $\R$ denotes the molecules' unperturbed end-to-end distance, is comprised of $61\,440$ polymers of $N=16$ segments. There are two hard walls at $x=0$ and $12\R$; one boundary attracts the A component of the blend whereas the other wall prefers the B component. The incompatibility toggles between $\chi N=4 \pm 4$ with a period of $T=800$ MC-steps
\footnote{Since the minimal value of the incompatibility is $\chi N=0$, corresponding to an ideal mixture, the period, $T$, cannot be made arbitrarily large, \ie, $\Lambda T\ll h(0)^{2}$, in order to retain a spatial modulation of $m$ at all times. We choose this large amplitude in order to increase the strength of the additional contribution to the interface potential.}.
Since we expect that the contribution of the oscillation to the interface potential increases like the square of the oscillation amplitude, we use a large variation of $\chi N$. The period, $T$, roughly corresponds to $0.42$ the time a polymer diffuses a distance $\R$ using Smart MC moves \cite{MullerSL} or approximately $10$ times the time $\tau_{w}$ a polymer diffuses the intrinsic width, $w_{\rm SSL}\equiv \frac{\R}{\sqrt{6\chi N}}$ \cite{Helfand75} of the interface at $\chi N = 4$. Further details about model and simulation technique are deferred to the SI.

\begin{figure}[t]
\includegraphics[width=0.9\columnwidth]{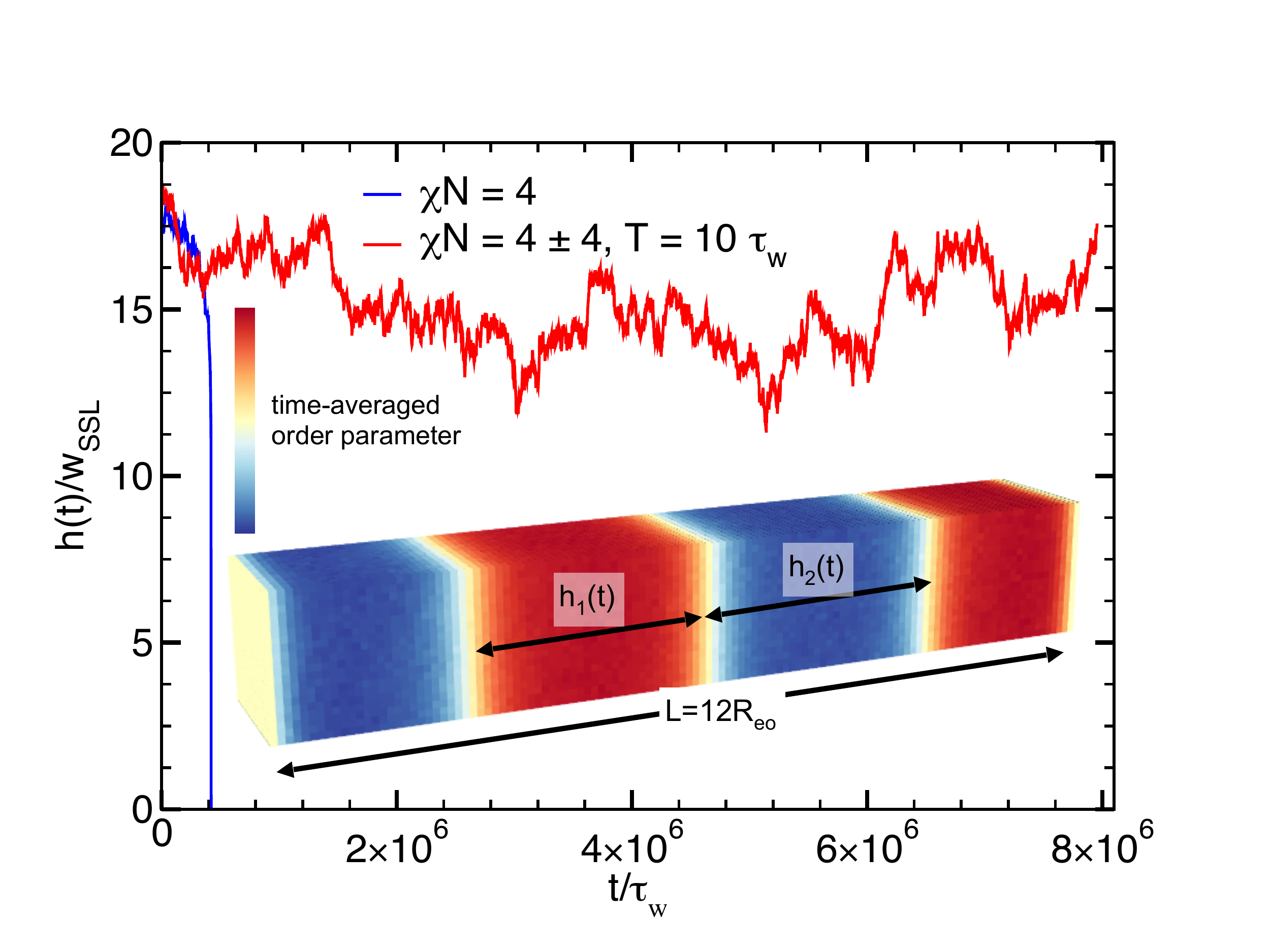}
\caption{$h(t)$ as observed by particle-based simulations for constant and oscillating incompatibility, $\overline{\chi N}=4$ and $\chi N= 4\pm4$, respectively. $h(t)=\frac{h_1(t)+h_2(t)}{2}$ is the average of the two interface distances present in the system. Lengths are measured in units of the strong-segregation approximation of the interface width, $w_{\rm SSL}=\R/\sqrt{6 \chi N}\approx 0.2\R$. Times are measured in units of $t_w = w_{\rm SSL}^2/D$, where $D$ denotes the self-diffusion coefficient in the disordered phase. The inset presents the profile, $m_0({\bf r})$, of the particle-based simulation averaged over the last $100T$.
}
\label{fig:4}
\end{figure}

\autoref{fig:4} shows the time evolution of the distance, $h(t)$, for the oscillating $\chi N$-parameter and compares the result to that of the equilibrium system with the time-averaged incompatibility $\overline{\chi N}=4$.  In the latter, equilibrium case, the interfaces approach each other and two interfaces annihilate after around $3\times10^7$ MC-steps or $4\times 10^5\tau_{w}$. In equilibrium there remains only a single interface that fluctuates around the middle of the film (soft-mode phase \cite{Parry90,Muller01e,Binder03}). In the driven system, where the incompatibility oscillates, there remain three interfaces in the system, and their mutual distances fluctuate around the preferred value, $h(\infty) \approx 3\R \approx 15 w_{\rm SSL}$ in the simulation that are extended to more than $6 \times 10^{8}$ MC-steps or $750\,000$ oscillation periods, $T$. The spacing of this lamellar structure of homopolymers, obtained by dissipative assembly, qualitatively agrees with the predictions of the Ginzburg-Landau model in \autoref{fig:2}. 

In summary we have analytically shown that temporal oscillations of the incompatibility between the species of a blend can qualitatively alter the interface potential from attractive to repulsive. The non-equilibrium phenomena occurs for moderate and slow oscillation periods, $T$, compared to the time the particles need to diffuse a distance comparable to the interfacial width. This dissipative assembly gives rise to preferred distances, $h(\infty)$, between planar interfaces, and this characteristic spacing of the lamellar morphology of the blend increases with the oscillation period like $h(\infty) \sim \sqrt{\Lambda T}$ for large $\Lambda T$, \ie, by varying the periodicity we can tailor the interface spacing without altering the molecular architecture or interactions. Whereas the preferred  distances, $h(\infty)$, are rather insensitive to the amplitude of the oscillations, we expect that the strength of the repulsion or the curvature of the interface potential around the preferred distance depends quadratically on the oscillation amplitude. Our particle-based simulations suggest that thin films of polymer blends are ideal model systems to validate our predictions because of the sensitivity of the phase behavior to minuscule changes of segmental interactions and the large time and length scales of polymer materials. 

Financial support from the Deutsche Forschungsgemeinschaft within the CRC 1073 / TPA03 is gratefully acknowledged. MM thanks the KITP Santa Barbara, supported by the National Science Foundation under Grant No. NSF PHY-1748958, for hospitality. We thank the von Neumann-Institute for Computing for access to the supercomputer JUWELS at J{\"u}lich Supercomputing Center. Additional computational resources at the HLRN Berlin/G{\"o}ttingen were made available. 

\clearpage
\newpage

\begin{widetext}
\section*{Supplementary Information}
\begin{center}
{\Large Interface repulsion and lamellar structures in thin films\\
  of homopolymer blends due to thermal oscillations}  \\[5mm]
  Louis Pigard and Marcus M{\"u}ller\\[5mm]
University of Goettingen, Institute for Theoretical Physics, 37077 G{\"o}ttingen, Germany
\end{center}
\end{widetext}
\subsection{A: Lyapunov functional, ${\cal L}_{2}$}
In this section we derive an approximation for the Lyapunov functional, ${\cal L}_{2}$ (cf.~\autoref{eqn:L2}), in two steps: First, we approximate \autoref{eqn:QPA} to obtain an estimate for $m_{1}[\bar m]$. To make progress, we approximate $\bar m^{2}\approx 1$ within the non-linear term $(3\bar m^2-1)m_1$, which is appropriate except for the interface region (\ie, we invoke a sharp-interface approximation for $\bar m(x)$ cf.~\autoref{fig:1}a and \autoref{eqn:si} below). The preferred distance, $h(\infty)$, however, is significantly larger than the interface width, $w$. This yields a linear relation between $m_{1}$ and $\bar m$
\begin{equation}
i\omega m_{1} = \nabla^{2} \left( \frac{2i}{\pi} \bar m +  2 m_{1} - \nabla^{2}m_{1}\right) 
\end{equation}
that can be solved by Fourier transform
\begin{equation}
\hat m_{1}(k) = \frac{-k^{2} \frac{2i}{\pi} \hat{\bar m}(k)}{i\omega+k^{2}(2+k^{2})} \equiv - \frac{\pi}{4} i k^{2}\hat A(k)  \hat{\bar m}(k)
\label{eqn:m1m}
\end{equation}
with the kernel
\begin{equation}
\hat A(k) =  \frac{8}{\pi^{2}}\frac{1}{i\omega+k^{2}(2+k^{2})} \,.
\end{equation}
Here and in the following, the hat indicates the Fourier transform 
\begin{equation}
\hat m_{1}(k)  = \frac{1}{\sqrt{2\pi}}\int \rd x\; m_{1}(x) e^{-ikx}\,.
\end{equation}

Second, we approximate \autoref{eqn:mu2} by neglecting the term $\bar m m_{1}^{2}$, which is justified because $|m_{1}|\ll 1$ as shown in \autoref{fig:3}. This results in the linear relation
\begin{equation}
\frac{\mu_{2}[\bar m]}{\kT {\cal N}} \approx  \frac{4}{\pi}\Im(m_{1}[\bar m])\,.
\end{equation}
Using \autoref{eqn:m1m} we obtain
\begin{eqnarray}
\frac{\hat \mu_{2}[\bar m](k)}{\kT {\cal N}} &\approx&\Im\left(-i k^{2}\hat A(k) \right) \hat{\bar m}(k) \\
&=& - k^{2} \Re\left(\hat A(k)\right) \hat{\bar m}(k)
\end{eqnarray}
because $\bar m$ is real and $\hat A$ is an even function of $k$. Since
$
\hat \mu_{2}(k) = \frac{\delta {\cal L}_{2}[\hat{\bar m}]}{\delta \hat {\bar m}(-k)}
$,
this chemical potential corresponds to the Lyapunov functional 
\begin{eqnarray}
\frac{{\cal L}_{2}[\bar m]}{\kT {\cal N}} 
&=& -\frac{1}{2}\int \rd k \; \Re\left(\hat A(k)\right) |ik\hat{\bar m}(k)|^{2} \\
&=&  -\frac{1}{2}\int \rd x \rd y \; \nabla_{x} \bar m(x) \; K(x-y) \; \nabla_{y} \bar m(y) \nonumber  \\
\mbox{with}\; K(z)&=&\frac{1}{2\pi} \int \rd k \; e^{ikz} \Re\left(\hat A(k)\right) \\
&=& \Re\left( \frac{1}{2\pi} \int \rd k \; e^{ikz} \hat A(k)\right)\\
&=&\frac{1}{\sqrt{2\pi}}\Re(A(z))
\end{eqnarray}
because $\hat A$ is even in $k$. 

\begin{comment}
{\color{gray}{
Attention: $\hat K_{\rm old}$ in the following paragraph corresponds to $\hat K_{\rm new}/\sqrt{2\pi}$ in the main text.
\begin{eqnarray}
\frac{\mu_{2}(x)}{\kT{\cal N}} &=& \frac{1}{\kT{\cal N}}\frac{\delta {\cal L}_{2}}{\delta \bar m(x)} \\
&=& \int \rd y\; \nabla_{x} K(x-y) \; \nabla_{y} \bar m(y) \nonumber \\
&=& \frac{1}{2\pi} \int \rd y \rd k\; \nabla_{x}  e^{ik(x-y)} \Re\left(\hat K(k)\right) \; \nabla_{y} \bar m(y) \nonumber \\
&=& \frac{1}{(2\pi)^{3/2}} \int \rd y \rd k \rd q\;  e^{ik(x-y)} ik \Re\left(\hat K(k)\right) \nabla_{y} \bar e^{iqy}\hat{\bar m}(q) \nonumber \\
&=& \frac{-1}{(2\pi)^{3/2}} \int \rd y \rd k \rd q\;  e^{ik(x-y)+iqy} k \Re\left(\hat K(k)\right) q \hat{\bar m}(q)\nonumber \\
&=& \frac{1}{\sqrt{2\pi}} \int \rd k \;  e^{ikx}  \; (-k^{2}) \Re\left(\hat K(k)\right) \hat{\bar m}(k)\nonumber \\
\frac{\hat \mu_{2}(k)}{\kT{\cal N}}  &=& -k^{2}  \Re\left(\hat K(k)\right) \hat{\bar m}(k)
\end{eqnarray}
}}
\end{comment}

In the following we use the sharp-interface approximation 
\begin{equation}
\nabla \bar m(x,t) \approx 2 \delta(x+h(t))-2\delta(x)+2\delta(x-h(t)) 
\label{eqn:si}
\end{equation}
which is approximate for $h(t)\gg w$. This sharp-interface approximation yields ${\cal F}[\bar m]\approx$ const and 
\begin{equation}
\frac{{\cal L}_{2}[\bar m]}{\kT {\cal N}} = -2 \left[ 3K(0)-4K(|h|)+2K(2|h|)\right]\,.
\end{equation}

To make progress, we decompose the kernel into
\begin{eqnarray}
\hat A(k) &=& \frac{4}{\pi^2}\left(\frac{1}{1-\lambda_{+}^{2}} \frac{1}{k^{2}+\lambda_{+}^{2}} +  \frac{1}{1-\lambda_{-}^{2}} \frac{1}{k^{2}+\lambda_{-}^{2}}\right) \nonumber \\
\mbox{with}\; \lambda_{\pm}^{2} &=& 1\pm\sqrt{1-i\omega} 
\end{eqnarray}
and obtain \autoref{eqn:res}
\begin{eqnarray*}
K(z) &=& \frac{2}{\pi^2}\Re \left( \frac{e^{-\lambda_{+}|z|}}{(1-\lambda_{+}^{2})\lambda_{+}} +  \frac{e^{-\lambda_{-}|z|}}{(1-\lambda_{-}^{2})\lambda_{-}}  \right)\,.
\end{eqnarray*}
In the limit of slow oscillations, $\omega \to 0$, we find $\lambda_{+}\approx2$ and $\lambda_{-}=\sqrt{i\omega/2}=(1+i)\sqrt{\omega}/2$.
Since the interface repulsion occurs for distances $h\gg w$, we can ignore the rapidly damped contribution with  $\lambda_{+}$ as well as the interaction between the two outer interfaces at $x=-h$ and $x=+h$. Thus, we approximate \autoref{eqn:Feff} by
\begin{eqnarray}
\frac{g(h)}{\kT{\cal N}} &=& \frac{F_{\rm eff}[h]}{2\kT{\cal N}} \approx 4 \epsilon^2 K(|h|) \\
&\approx& 4 \epsilon^{2} \Re \left(  \frac{2 e^{-(1+i)\sqrt{\omega}|h|/2}}{\pi^{2}(1+i)\sqrt{\omega}} \right) \\
&=& \frac{4\epsilon^{2}e^{-\eta} }{\pi^{2}\sqrt{\omega}} \left[ \cos\left(\eta\right)-\sin\left(\eta\right)\right] 
\;\mbox{with} \;\eta = \frac{\sqrt{\omega}|h|}2 \nonumber
\end{eqnarray}
This approximation of the interface potential exhibits minima at $h_{n}(\infty)=\frac{\pi}{\sqrt{\omega}}(1+4n)=\sqrt{\frac{\pi \Lambda T}2}(1+4n)$ with $n\in \mathbb{N}$.

\subsection{B: Energy dissipation}

\begin{figure}[t]
\includegraphics[width=0.9\columnwidth]{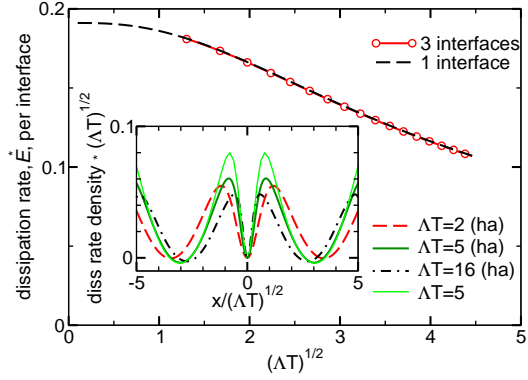}
\caption{Energy dissipation, $\dot{\cal E}_\textrm{ha}$, per interface obtained within the harmonic ansatz as a function of the square root of the oscillation period, $\sqrt{\Lambda T}$, for $\epsilon=0.5$. Both, data for systems that contain $3$ interfaces as well as for systems that contain only a single interface are included. 
The inset presents the spatial density of the dissipation rate normalized by $\sqrt{\Lambda T}$, \ie, $- 4 \epsilon m_0(x) m''(x)\sqrt{\frac{\Lambda}{T}}$, as a function of $x/\sqrt{\Lambda T}$ for $\Lambda T=2$, 5 and 16. The inset also shows $- \epsilon\sqrt{\frac{\Lambda}{T}}\left[m^{2}(x,0^-)-m^{2}(x,{T}/{2}^-)\right]$ for $\Lambda T=5$.
}
\label{fig:5}
\end{figure}

The instantaneous change in the energy of the system is the sum of the work performed on the system by changing the incompatibility, $\alpha$, which quantifies the pairwise repulsion between the different homopolymer segments, and the energy dissipated by the system into the environment. In a time-periodic nonequilibrium steady state, the change in total energy vanishes during one period. Therefore, in analogy to the seminal work of Tagliazucchi, Weiss, and Szleifer \cite{Igal1},  we define the average dissipation rate, $\dot{\cal E}$, by 

\begin{eqnarray}
\dot{\cal E} &\equiv& - \frac{1}{T} \int_0^T {\rm d}t\; \int {\rm d}x \; \left[\frac{1}{2} \frac{{\rm d}\alpha}{{\rm d}t}m^{2}(x,t)\right] \\
&=& - \frac{\epsilon}{T}    \int {\rm d}x \; \left[m^{2}(x,0^-)-m^{2}(x,{T}/{2}^-)\right] \label{eqn:sim}
\end{eqnarray}
where the order parameter is evaluated before the toggling of the interactions at $t=0$ and $T/2$. Within the harmonic ansatz, Eq.~(3), we obtain the simple expression
\begin{eqnarray}
\dot{\cal E}_\textrm{ha} &=& - \frac{4 \epsilon^2}{\pi}  \omega \Lambda \int {\rm d}x\; m_0(x) \Re\left(m_1(x)\right)  \\
&=&  - \frac{4 \epsilon}{T}  \int {\rm d}x\; m_0(x) m''(x)
\end{eqnarray}  
where $m''(x)$ denotes the out-of-phase response of $m$ that is presented in \autoref{fig:3}.
    
\autoref{fig:5} presents the dissipation rate as a function of oscillation period, $T$, according to \autoref{eqn:sim}, setting ${\Lambda}=1$. The dissipation rate gradually decreases with the oscillation period. In agreement with Ref.~\cite{Igal1} the energy dissipation \textit{per oscillation cycle}, $\dot{\cal E}T$, vanishes in the limit $T\to 0$ because the order parameter cannot respond to the fast oscillation, \ie, $m^{2}(x,{T}/{2}^-) \xrightarrow{T\rightarrow 0} m^{2}(x,0^-)$ or, equivalently, the out-of-phase response, $m''$ vanishes.

At the oscillation period, $\Lambda T^* \approx 1.7$ (for $\epsilon=0.5$), the number of interfaces changes from $1$ for $T<T^*$ to $3$ for $T>T^*$, in accord with \autoref{fig:2}b. The increase of the dissipation rate also is approximately three-fold. Therefore we plot in \autoref{fig:5} the dissipation rate per interface and the data for systems with $1$ and $3$ interfaces collapse.

This suggests that the dissipation is localized around the interfaces, and we display in the inset of \autoref{fig:5} the spatial density of the dissipation rate, $-\frac{\epsilon}{T}\left[m^{2}(x,0^-)-m^{2}(x,{T}/{2}^-)\right]$, or the corresponding expression, $-\frac{4\epsilon}{T}m_0m''$, within the harmonic ansatz. The dissipation extends a distance proportional to $\sqrt{T}$ away from each interface.  We have rescaled the spatial distance $x$ from the interface by a factor $1/\sqrt{T}$ and normalized the dissipation rate density by $\sqrt{T}$ to highlight this dependence by the collapse of the data for different $T$. This spatial scaling agrees with the dependence of the preferred distance between interfaces in the lamellar structure produced by the dissipative assembly.

In the limit of extremely long periods, $T\to \infty$, the instantaneous order-parameter profile corresponds to the equilibrium profile for the instantaneous value, $\alpha(t)$, of the incompatibility. Thus, in accord with Ref.~\cite{Igal1}, the out-of-phase response vanishes and so does the dissipated energy, $\dot{\cal E}T$, within an oscillation cycle. We do not observe this limiting behavior in \autoref{fig:5}. The equilibration times of the system with a typical interface distance studied here are on the order $\Lambda T_\textrm{eq} \sim {\cal O}(10^4)$ (see Figures.~\ref{fig:1}b and \ref{fig:2}a) whereas the largest oscillation periods in \autoref{fig:5} is only $\Lambda T = 20$. Thus the limit $T\gg T_\textrm{eq}$ is not reached for the data presented in \autoref{fig:5}.

\subsection{C: Simulation model and technique}
For the particle-based simulations we use a soft, coarse-grained model \cite{mm11b} of a symmetric binary polymer blend. Polymers are represented by a bead-spring model, where neighboring beads along the linear, flexible molecular contour are connected by harmonic springs. The bonded interactions take the form
\begin{equation}
\frac{{\cal H}_{\rm b}(\{\textbf{{R}}_{i,j}\})}{k_{\rm B}T}=\frac{3(N-1)}{2\R^{2}}\sum^{n}_{i=1}\sum^{N}_{j=2}(\textbf{{R}}_{i,j}-\textbf{{R}}_{i,j-1})^{2}\,,
\end{equation}
where $\textbf{{R}}_{i,j}$ denotes the spatial position of the coarse-grained segment, $1\leq j\leq N$, on polymer, $1\leq i\leq n$. $k_{\rm B}$ is the Boltzmann constant, while $T$ denotes the temperature. We choose the chain discretization $N=16$ and a large invariant degree of polymerization, $\sqrt{\Nbar}=\rho_0 \R^3/N=1280$. The symmetric mixture is comprised of the same number of A and B polymers, and both species occupy the same segmental volume, $1/\rho_0$, and have the same unperturbed end-to-end distance, $\R$.

The nonbonded interactions, ${\cal H}_{\rm nb}$, represent the small compressibility of the binary polymer melt and the incompatibility between A and B species. 
\begin{eqnarray}
\frac{{\cal H}_{\rm nb}(\{\textbf{{R}}_{i,j}\})}{k_{\rm B}T\sqrt{\Nbar}}&=&\int \frac{{\rm d}\textbf{r}}{\R^3}\left\{\frac{\kappa N}{2}[\hat{\phi}_{A}+\hat{\phi}_{B}-1]^{2}\right.\nonumber\\
&&\hspace*{1.4cm}\left.-\frac{\chi N}{4}[\hat{\phi}_{A}-\hat{\phi}_{B}]^{2}\right\}\,. \label{eqn:Hnb}
\end{eqnarray}
where the A volume fraction is computed from the particle positions according to
\begin{equation}
\hat{\phi}_{A}(\textbf{r}|\{\textbf{{R}}_{i,j}\})=\frac{1}{\rho_{0}}\sum^{n_A}_{i=1}\sum^{N}_{j=1}\delta(\textbf{r}-\textbf{{R}}_{i,j})
\end{equation}
where the sum runs over all $n_A$ A polymers in the system. A similar relation holds for the volume fraction of B. We use $\kappa N=16$ to restrain fluctuations of the local density and the incompatibility parameter $\chi N$ toggles between $0$ and $8$. 

The nonbonded interactions are evaluated on a collocation grid with a spacing $\Delta L=0.1\R$, \ie, for each point,${\bf c}$, on the collocation grid, we define 
\begin{equation}
\hat{\phi}_{A}(\bc) = \int \frac{{\rm d}^{3}\br}{\Delta L^{3}}\;\Pi(\bc,\br) \hat \phi_{A}(\br) = \frac{1}{\rho_{\rm o}\Delta L^{3}} \sum_{i=1}^{n_A} \sum_{j=1}^{N} \Pi({\bf c},\textbf{{R}}_{i,j})\,.
\end{equation}
The assignment function, $\Pi$, is $1$ for all points, $\br$, within a volume $\Delta L^3$ around ${\bf c}$, and vanishes otherwise. We emphasize that the quadratic nonbonded interactions in \autoref{eqn:Hnb} are equivalent to pairwise interactions between particles \cite{mm11b}. Thus, a periodic variation of $\chi N$ with time is equivalent to a periodic variation of the interparticle potential. The period-averaged interaction potential \cite{Igal1,Igal2} exactly corresponds to the interactions in a system with constant $\overline{\chi N}=4$ and the concomitant free-energy functional is $\bar{\cal F}$, \ie, the leading-order term in ${\cal F}_{\rm eff}$ that is independent from the amplitude $\epsilon$.

The mixture is confined into a thin film that is bounded by impenetrable surfaces at $x=0$ and $x=12\R$. Periodic boundary conditions are applied in the two lateral directions.

The surfaces of the film attract the different species of the blends with opposite and equal interactions. An A segment within the distance $\Delta L$ from the right wall decreases the energy by $H=2 \kT$ and a B segment increases the energy by the same amount. The surface free-energy difference of the A-rich phase and the B-rich phase of the symmetric blend is approximately $\frac{\Delta \gamma \R^2}{\kT\sqrt{\Nbar}}\approx 2 H N \Delta L/\R$ \cite{Muller98b}. The preferred phase wets the surface if this value is larger than the interface tension \cite{Young}. In this case, the preferred species forms a layer at the wall and the AB interface is repelled by the surface in order to form a macroscopically thick surface layer of the preferred phase \cite{schick1990introduction}. Using the analytical prediction for the interface tension, $\frac{\gamma_{\rm SSL} \R^2}{\kT\sqrt{\Nbar}}=\sqrt{\chi N/6}$ \cite{Helfand75}, we obtain $\Delta \gamma/\gamma=\frac{2\sqrt{6}HN\Delta L}{\sqrt{\chi N}\R} = 7.84>1$ in our simulations. Since the A-rich phase wets the left wall and the $B$-rich phase wets the right surface, there is a single AB interfaces in equilibrium, that runs parallel to the surfaces and fluctuates around the middle of the film -- soft-mode phase \cite{Parry90,Muller01e,Binder03}.

We use the single-chain-in-mean-field algorithm \cite{Daoulas06b} to exploit the separation between the strong bonded and weak non-bonded interactions.

Polymer conformations are updated by smart Monte-Carlo moves giving rise to Rouse-like dynamics \cite{MullerSL}. In the disordered phase, a polymer needs about $\tau = \R^2/D \approx 1900$ MC-steps to diffuse its mean-squared extension, $\R^2$, in the three-dimensional simulations, or about $80$ MC-steps to diffuse the distance $w_{\rm SSL}^2$ The large-scale simulations use the implementation SOMA \cite{SOMA} on multiple GPUs.

%The initial morphology with three interfaces can be produced in two steps, starting from the disordered phase $m=0$. Initially, the preference of the film surfaces will give rise to the formation of enrichment layers of A and B at the respective surfaces. These composition variations will be exponentially amplified in time by the spinodal instability of the disordered phase. The dominant mode with the highest growth rate is $k^2=\frac{\alpha}{2\gamma^2}$ where $\alpha=\frac{\chi N-2}{2}$ and $\gamma^2=\frac{\R^2}{18}$, appropriate for weak segregation. In order to introduce three interfaces we desire $k=\frac{3\pi}{L}$. We can tune $\chi N$ and $L$ in order to create the appropriate mode. However, this results in $\chi N-2=\frac{2}{9}\left(\frac{3\pi \R^2}{L}\right)^2\approx 0.14$. This is very close to the critical point which means thermal fluctuations perturb the mean-field predictions significantly. 

%In this sense, it is better to increase the extension of the chains $\R^2$ from $1$ to $\R^2=3(\chi N-2)\left(\frac{L}{3\pi}\right)^2=9.7$ at $\chi N=4$ (or take a compromise between the two options). The growth will eventually stop after several characteristic growth times $\tau_\mathrm{g}=\frac{8\gamma^2}{\alpha^2D}=1000$ Monte Carlo steps (we choose 10$\tau_\mathrm{g}$). Finally, we let the chains shrink to the original extension $\R^2=1$ and equilibrate the width of the interface for $10\tau_\mathrm{g}$.

\bibliography{bibtex.bib}
\bibliographystyle{apsrev}

\end{document}